\documentclass[
]{elsarticle}

\usepackage{graphicx}
\usepackage{amssymb}   
\usepackage{amsfonts}
\usepackage{amsmath}
\usepackage{color}
\usepackage{natbib}
\usepackage{lineno,hyperref}
\modulolinenumbers[5]

\journal{Journal of Crystal Growth}

\begin{document}

\begin{frontmatter}

\title{Nominal vs. actual supersaturation of solutions}

\author{Alexander~Borisenko}
\address{National Science Center "Kharkiv Institute of Physics and Technology", Akademichna Street 1, 61108 Kharkiv, Ukraine}
\ead{borisenko@kipt.kharkov.ua}




\begin{abstract}
Following the formalism of the Classical Nucleation Theory beyond the dilute solution approximation, this paper considers a difference between the actual solute supersaturation (given by the present-to-saturated solute activity ratio) and the nominal supersaturation (given by the present-to-saturated solute concentration ratio) due to formation of subcritical transient solute clusters, called heterophase fluctuations.
Based on their distribution function, we introduce an algebraic equation of supersaturation that couples the nominal supersaturation of a binary metastable solution with its actual supersaturation and a function of the specific interface energy and temperature. The applicability of this approach is validated by comparison to simulation data [E. Clouet et al., Phys. Rev. B \textbf{69}, 064109 (2004)] on nucleation of Al$_{3}$Zr and Al$_{3}$Sc in model binary Al alloys.
\end{abstract}


\end{frontmatter}


\section{\label{Introduction}Introduction}

Because decomposition of supersaturated solutions occurs in many natural phenomena and important technological processes, this problem  traditionally attracts much attention (for some recent results see, e.g., Refs.~\cite{Lin,Dhivya,Binder,Greer,Warrier,Bi,Lifanov,Legg,Peters,Mangere} and references therein).

From the thermodynamic point of view, the solution properties, including the nucleation driving force (see Eq.~(\ref{d_mu}) below), are determined by the \textit{actual supersaturation}: 
\begin{eqnarray} \label{S_act}
S_{\rm{act}}=a / a^{\rm{sat}}, 
\end{eqnarray}
where $a$ is the solute thermodynamic activity and $a^{\rm{sat}}$ corresponds to the solubility (saturation) limit. 

On the other hand, the only experimentally available value is the \textit{nominal  supersaturation}: 
\begin{eqnarray} \label{S_nom}
S_{\rm{nom}}=c_{\rm{tot}}/c_{\rm{tot}}^{\rm{sat}}, 
\end{eqnarray}
where $c_{\rm{tot}}$ is the total solute concentration, calculated as the total number of solute atoms $N$, divided by the solution volume $V$: 
\begin{eqnarray} \label{c}
c_{\rm{tot}}=N/V,
\end{eqnarray}
and $c_{\rm{tot}}^{\rm{sat}}$ corresponds to the solubility (saturation) limit.

A very popular dilute solution (DS) model (see, e.g., Ref.~\cite{Landau}) does not distinguish between the nominal supersaturation and the actual supersaturation: $S=S_{\rm{nom}}=S_{\rm{act}}$ (cf., e.g., Eqs.~(2.13) and (2.14) of Kashchiev~\cite{Kashchiev}), which appears to be a good approximation for essentially weak solutions, where the vast majority of solute atoms is in the monomer state. 

According to the Frenkel's concept~\cite{Frenkel} of heterophase fluctuations (HF), pretransition processes near the saturation limit lead to formation of dimers, trimers and larger transient solute clusters.

Recent studies demonstrate, that HF in solutions play an important role in the nucleation kinetics~\cite{B_T,Lepinoux2009,Lepinoux2010,S_O,Borisenko1,Borisenko2}. Here we reveal a crucial role of HF in the thermodynamic treatment of supersaturated metastable solutions.

Based on the distribution function of HF, in Section~\ref{EqOfSs} we derive an algebraic \textit{equation of supersaturation} of a binary metastable solution, coupling the values of the nominal~(\ref{S_nom}) and the actual~(\ref{S_act}) supersaturation with a function of the specific cluster-solution interface energy $\sigma$ and temperature $T$ (see Eq.~(\ref{EOS}) below). In Section~\ref{sec3} we apply this equation to compare its results to some simulation data~\cite{Clouet} on nucleation of Al$_{3}$Zr and Al$_{3}$Sc in model binary Al alloys (see Figs.~\ref{Fig:2}--\ref{Fig:4} below) and demonstrate that a good agreement is achieved in some cases and in all cases Eq.~(\ref{EOS}) gives a much better agreement than the dilute solution model does. We leave Section~\ref{Conclusions} for conclusions and outlook of future tasks.

\section{\label{EqOfSs}Equation of supersaturation of a binary metastable solution}

In this paper we consider solutions with a single type of solute molecules which do not dissociate.

Within the Classical Nucleation Theory (CNT), the Gibbs free energy change on forming a  cluster of $n$ solute monomers (at constant pressure and temperature) 
is (see, e.g., Eq.~(3.39) of Kashchiev~\cite{Kashchiev}): 
\begin{eqnarray} \label{dG}
\Delta G\left(n\right)= \Delta \mu \cdot n + k_{\rm{B}}T \cdot \alpha_{n} \cdot n^{\frac{2}{3}}.
\end{eqnarray}
The first term in the right-hand side of Eq.~(\ref{dG}) contains the chemical potential change on clusterization, equal to the nucleation driving force taken with the opposite sign (see, e.g., Eq.~(2.13) of Kashchiev~\cite{Kashchiev}): 
\begin{eqnarray} \label{d_mu}
\Delta \mu= - k_{\rm{B}}T \cdot \ln \left( S_{\rm{act}} \right), 
\end{eqnarray}
where $k_{\rm{B}}$ is the Boltzmann's constant. The value given by Eq.~(\ref{d_mu}) can change sign depending on the value of $S_{\rm{act}}$.
The second term in the right-hand side of Eq.~(\ref{dG}) is the free energy of the cluster-solution interface. 
For a spherical cluster, the dimensionless size-dependent specific interface energy $\alpha_{n}$ is 
\begin{eqnarray} \label{alpha}
\alpha_{n}= 3\cdot \sqrt[3]{4\pi \omega^{2}/3}\cdot \sigma_{n} /k_{\rm{B}}T, 
\end{eqnarray}
where $\omega$ is a volume per solute monomer in the cluster and $\sigma_{n}$ is a size-dependent specific cluster-solution interface energy. The value given by Eq.~(\ref{alpha}) is positively defined. 


By utilizing Eqs.~(\ref{dG}) and~(\ref{d_mu}), the equilibrium distribution function of HF of different sizes, introduced by Frenkel~\cite{Frenkel}, can be presented as follows (see, e.g., Eq.~(7.17) of Kashchiev~\cite{Kashchiev}): 
\begin{eqnarray} \label{Frenkel}
c_{n}^{0}=c_{1}^{\rm{sat}}\cdot \exp\left[\ln \left( S_{\rm{act}} \right) \cdot n - \alpha_{n} \cdot \left(n^{\frac{2}{3}} - 1 \right)\right],
\end{eqnarray}
where $c_{1}^{\rm{sat}}$ is a saturated concentration of solute monomers.
Eq.~(\ref{Frenkel}) is a particular form of the Boltzmann distribution for the clusters with the energy spectrum~(\ref{dG}), 
specially normalized to give the concentration (number density) of clusters of size $n$.

For $n=1$, to get a trivial identity $c_{1}^{0}=c_{1}$ from Eq.~(\ref{Frenkel}), one has to set 
\begin{eqnarray} \label{a_c}
S_{\rm{act}}=a / a^{\rm{sat}}=c_{1} / c_{1}^{\rm{sat}}.
\end{eqnarray}
Eq.~(\ref{a_c}) is a direct result of Eq.~(\ref{Frenkel}) and, therefore, originates from the choice of the work of clusterization in the form~(\ref{dG}) and~(\ref{d_mu}). It means that, within CNT, the solute thermodynamic activity is determined by the concentration of solute monomers only. This approximation seems natural, provided that the diffusivity of solute monomers greatly exceeds that of solute clusters. 

From Eq.~(\ref{Frenkel}) one can see that, in the supersaturated case $S_{\rm{act}} > 1$, the equilibrium distribution of clusters diverges rapidly as $n \to \infty$. To avoid this unphysical behavior, one has to take into account a nonzero value of the net cluster flux along the size axis: 
\begin{eqnarray} \label{J}
J_{n, n+1}\left(t\right)= w^{(+)}_{n, n+1} c_{n}\left(t\right) - w^{(-)}_{n+1, n} c_{n+1}\left(t\right), 
\end{eqnarray}
where
$w^{(+)}_{n, n+1}$ and 
$w^{(-)}_{n+1, n}$ are, respectively, the rates of attachment 
and detachment of solute monomers at the cluster-solution 
interface. 

The special case, when the net cluster flux~(\ref{J}) is zero for any $n$, corresponds to the state of detailed balance, when the HF distribution function takes its equilibrium form~(\ref{Frenkel}).

In the steady-state nucleation regime, the net cluster flux~(\ref{J}) is assumed to be a step function of size: 
\begin{eqnarray}
J_{n, n+1}^{\rm{st}}= \nonumber
\begin{cases}
J,     1 \leq n \leq n_{\rm{max}}; \\ 
0,     n > n_{\rm{max}},
\end{cases}
\end{eqnarray}
where the steady-state nucleation rate is (see, e.g., Eq.~(13.30) of Kashchiev~\cite{Kashchiev}): 
\begin{eqnarray} \label{J_nucl}
J=\left\{\sum_{n=1}^{n_{\rm{max}}} \left[w^{(+)}_{n, n+1}c_{n}^{0}\right]^{-1}\right\}^{-1},
\end{eqnarray}
and the rate of attachment of solute monomers at the interface in the diffusion-limited case is (see, e.g., Eq.~(10.23) of Kashchiev~\cite{Kashchiev}):
\begin{eqnarray}\label{w+}
w^{(+)}_{n, n+1}=4\pi \sqrt[3]{3\omega/ 4\pi}\left(1+n^{\frac{1}{3}}\right)\cdot D\left(1+n^{-\frac{1}{3}}\right) \cdot c_{1} , 
\end{eqnarray}
$D$ being the solute diffusion coefficient.

In the steady-state nucleation regime, the HF distribution function is (see, e.g., Eq.~(13.17) of Kashchiev~\cite{Kashchiev}):
\begin{eqnarray} \label{g_J_a}
c_{n}^{J}=c_{n}^{0}\cdot J \cdot \sum_{m=n}^{n_{\rm{max}}} \left[w^{(+)}_{m, m+1}c_{m}^{0}\right]^{-1}
\end{eqnarray}
for $1 \leq n \leq n_{\rm{max}}$ and $c_{n}^{J}=0$ for $n > n_{\rm{max}}$.

In the undersaturated and saturated solutions $S_{\rm{act}} \leq 1$ and from Eq.~(\ref{Frenkel}) one gets  $\lim_{n \to \infty} c_{n}^{0}=0$ and, therefore, formally setting $n_{\rm{max}} \to \infty$, from Eqs.~(\ref{J_nucl}) and~(\ref{g_J_a}) one gets $J=0$ and $c_{n}^{J}=c_{n}^{0}$.
 
With the above speculations in mind, one can represent the total solute concentration~(\ref{c}) as a sum of HF contributions: 
\begin{eqnarray} \label{c_nom}
c_{\rm{tot}}=\sum_{n=1}^{n_{\rm{max}}} n \cdot c_{n}^{J}.
\end{eqnarray}
Eq.~(\ref{c_nom}) is also valid for the undersaturated and saturated cases, where $c_{n}^{J}=c_{n}^{0}$ and $n_{\rm{max}} \to \infty$. 

The saturated monomer concentration $c_{1}^{\rm{sat}}$ can be extracted from the total solubility $c_{\rm{tot}}^{\rm{sat}}$, using Eqs.~(\ref{c_nom}) and~(\ref{Frenkel}), as follows: 
\begin{eqnarray} \label{c_1}
c_{1}^{\rm{sat}}=c_{\rm{tot}}^{\rm{sat}}\Bigg{/}\sum_{n=1}^{\infty} n \cdot \exp\left[ - \alpha_{n} \cdot \left(n^{\frac{2}{3}} - 1 \right)\right].
\end{eqnarray}

With Eqs.~(\ref{c_nom}), (\ref{g_J_a}), (\ref{J_nucl}), (\ref{w+}) and~(\ref{Frenkel}) in mind, after some algebra one can express the nominal supersaturation~(\ref{S_nom}) as follows: 
%
\begin{eqnarray} \label{EOS}
S_{\rm{nom}}= \\  \nonumber
\frac{\sum\limits_{n=1}^{n_{\rm{max}}} \left \{n \cdot S_{\rm{act}}^{n} \cdot \exp\left[- \alpha_{n} \cdot \left(n^{\frac{2}{3}} - 1 \right)\right] \cdot 
\sum\limits_{m=n}^{n_{\rm{max}}} \frac{\exp\left[ \alpha_{m} \cdot \left(m^{\frac{2}{3}} - 1 \right)\right]}{\left(1+m^{-\frac{1}{3}}\right)\left(1+m^{\frac{1}{3}}\right)\cdot S_{\rm{act}}^{m} }\right \}}
{\left\{\sum\limits_{n=1}^{n_{\rm{max}}} \frac{\exp\left[\alpha_{n} \cdot \left(n^{\frac{2}{3}} - 1 \right)\right]}{\left(1+n^{-\frac{1}{3}}\right)\left(1+n^{\frac{1}{3}}\right)\cdot S_{\rm{act}}^{n}} \right\} 
\left\{\sum\limits_{n=1}^{\infty} n \cdot \exp\left[ - \alpha_{n} \cdot \left(n^{\frac{2}{3}} - 1 \right)\right]\right\}}.
\end{eqnarray}
%

Eq.~(\ref{EOS}) can be called an equation of supersaturation (ES) of a binary metastable solution, because it couples the nominal solute supersaturation $S_{\rm{nom}}$ with the actual supersaturation $S_{\rm{act}}$ and a function $\{\alpha_{n}\}$ of the specific cluster-solution interface energy and temperature. It is also valid for undersaturated and saturated thermodynamically equilibrium solutions, where one has to assume formally $n_{\rm{max}} \to \infty$.

\section{\label{sec3}Results and discussion}

To obtain a practically valuable result, ES~(\ref{EOS}) can be resolved to find the actual supersaturation $S_{\rm{act}}$ (or the thermodynamic driving force $k_{\rm{B}}T \cdot \ln \left( S_{\rm{act}} \right)$) as a function of  the nominal supersaturation $S_{\rm{nom}}$ and $\{\alpha_{n}\}$. From ES~(\ref{EOS}) one can see that the difference between the actual supersaturation and the nominal supersaturation becomes sizeable when HF are energetically cheap, i.e. when either $S_{\rm{act}}$ is large or $\{\alpha_{n}\}$ is small. 
On the other hand, for small $S_{\rm{act}}$ or large $\{\alpha_{n}\}$, when HF are energetically expensive, one can retain only the first terms in the sums in both the numerator and the denominator of ES~(\ref{EOS}) to obtain the DS result $S_{\rm{nom}}=S_{\rm{act}}$.

\begin{figure}
\centerline{\includegraphics
[width=0.7\textwidth]
{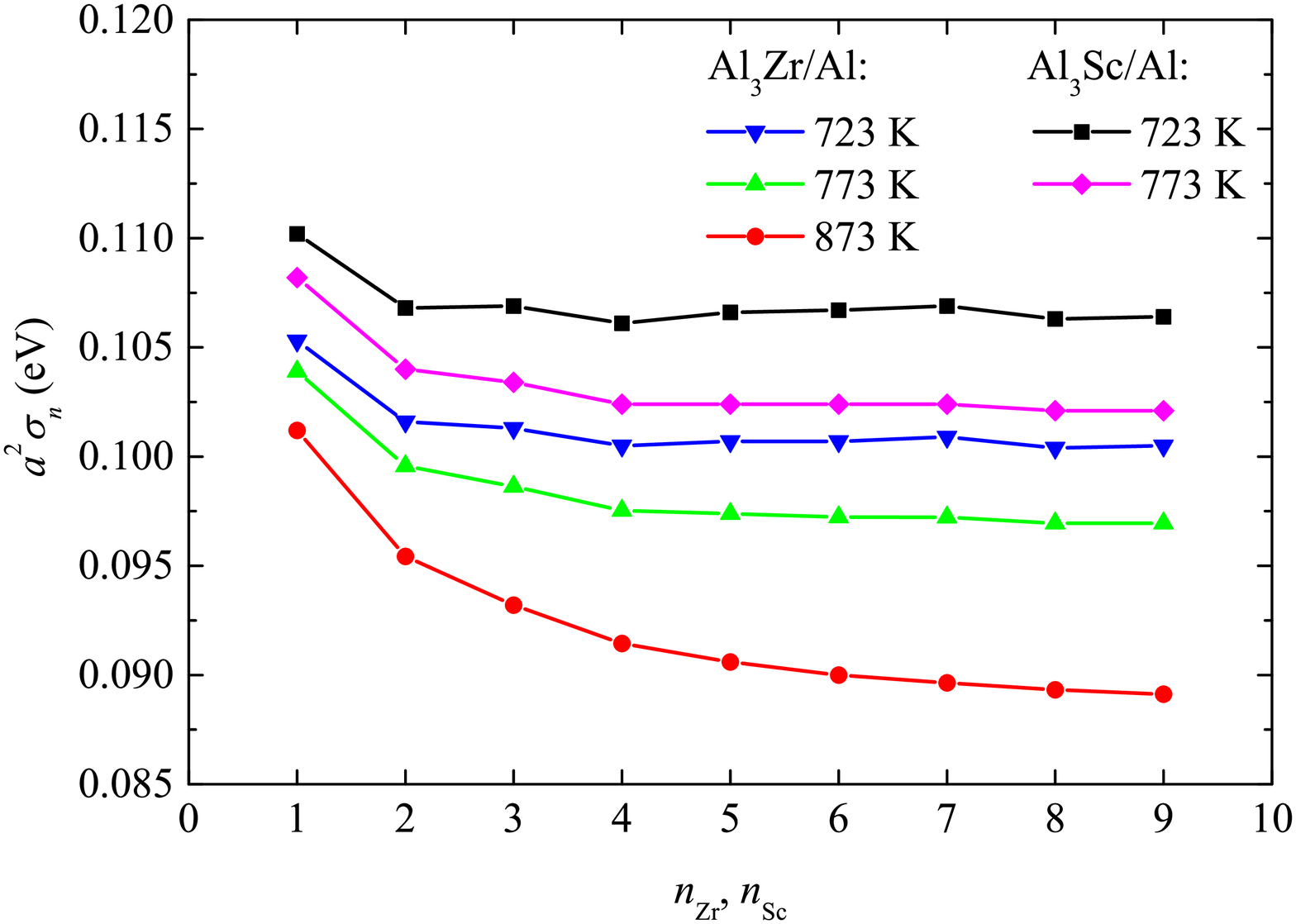}}
\caption{\label{Fig:1} Size-dependent specific interface energies $a^{2}\sigma_{n}$, calculated for indicated temperatures and interfaces from Eq.~(20) and Table~III of Ref.~\cite{Clouet} for $1 \leq n \leq 9$. For $n>9$ we take $\sigma_{n}=\sigma_{9}$. The lines are only guides for an eye.}
\end{figure}

\begin{figure}
\centerline{\includegraphics
[width=0.7\textwidth]
{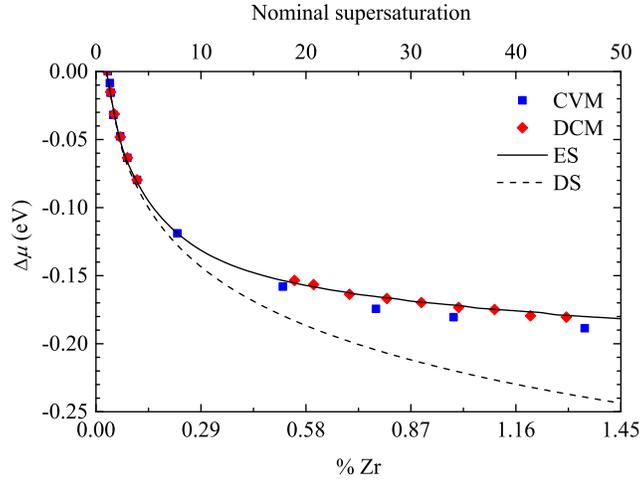}}
\caption{\label{Fig:2} The chemical potential change on clusterization (per Al$_{3}$Zr formula unit) as a function of the Zr volume fraction (bottom axis) or the nominal supersaturation (top axis) in the Al-Zr alloy at $T=723~{\rm{K}}$, obtained by CVM (squares) and DCM (diamonds) from the MC simulations~\cite{Clouet} and calculated (solid line) from Eq.~(\ref{d_mu}), with $S_{\rm{nom}}$ converted to $S_{\rm{act}}$ by ES~(\ref{EOS}). The  DS result $\Delta \mu= - k_{\rm{B}}T \cdot \ln \left( S_{\rm{nom}} \right)$ is indicated with the dashed line as a reference.}
\end{figure}

\begin{figure*}
\centerline{\includegraphics
[width=1.0\textwidth]
{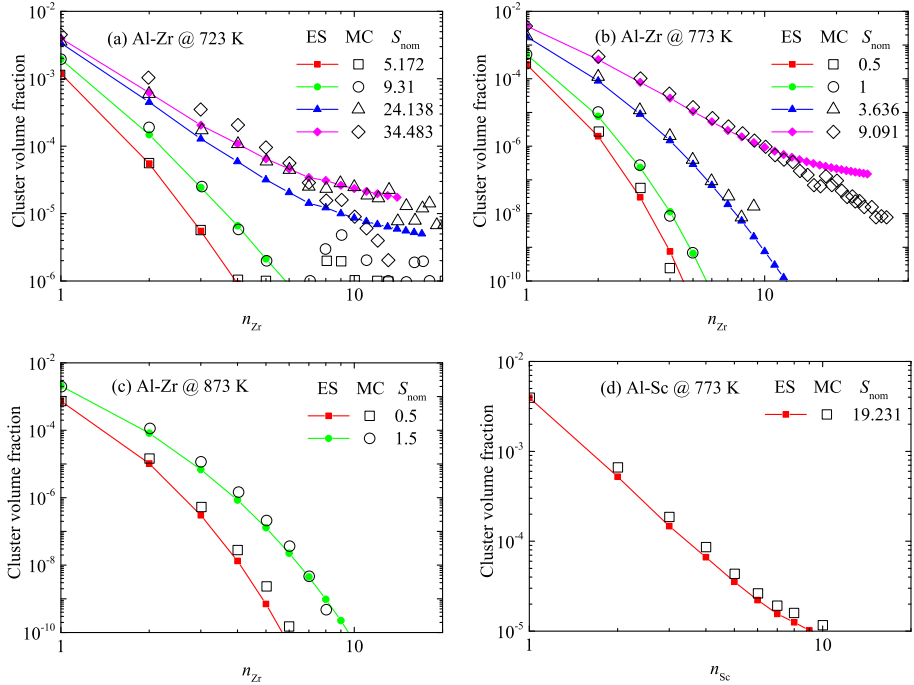}}
\caption{\label{Fig:3} Distribution functions for the Al$_{3}$Zr (a - c) and Al$_{3}$Sc (d) clusters, obtained from the MC simulations~\cite{Clouet} (large open symbols) and calculated (small filled connected symbols) from Eq.~(\ref{g_J_a}), with $S_{\rm{nom}}$ converted to $S_{\rm{act}}$ by ES~(\ref{EOS}). Temperatures and nominal supersaturations are as indicated. The lines are only guides for an eye.}
\end{figure*}

\begin{figure*}
\centerline{\includegraphics
[width=1.0\textwidth]
{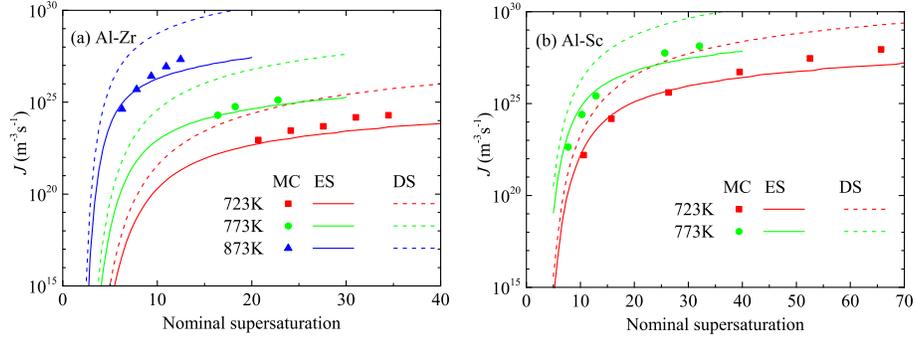}}
\caption{\label{Fig:4} Nucleation rates for the Al$_{3}$Zr (a) and Al$_{3}$Sc (b) clusters, obtained from the MC simulations~\cite{Clouet} (symbols) and calculated (solid lines) from Eq.~(\ref{J_nucl}), with $S_{\rm{nom}}$ converted to $S_{\rm{act}}$ by ES~(\ref{EOS}). Temperatures are as indicated. The dashed lines show the DS result with $S_{\rm{act}}=S_{\rm{nom}}$.}
\end{figure*}

To illustrate the practical applicability of ES~(\ref{EOS}), below we compare the results of the present approach to the data of the Monte Carlo (MC) simulations~\cite{Clouet} of the model Al-Zr and Al-Sc alloys, which exhibit homogeneous nucleation of the Al$_{3}$Zr and Al$_{3}$Sc phases, respectively. In Figs.~\ref{Fig:2}--\ref{Fig:4} below we demonstrate the simulated data together with the calculated ones, using size-dependent specific interface energies $\sigma_{n}$ shown in Fig.~\ref{Fig:1} and calculated from Eq.~(20) and Table~III of Ref.~\cite{Clouet} for $1 \leq n \leq 9$. For $n>9$ we take $\sigma_{n}=\sigma_{9}$. 
For the face-centered cubic lattice used in the MC simulations~\cite{Clouet}, we calculate the volume per Al$_{3}$Zr or Al$_{3}$Sc formula unit ($4$ lattice sites) as $\omega=4 \cdot a^{3}/4=a^{3}$, taking the lattice constant to be equal to that of the aluminium $a=4.05\cdot10^{-10}~{\rm{m}}$. The values of the upper summation limit $n_{\rm{max}}$ are chosen to satisfy the condition $c_{n_{\rm{max}}}^{J}\big{/}c_{n_{\rm{max}}}^{0} \leq 10^{-2}$. The values of other parameters are adopted from Ref.~\cite{Clouet} and collected in Table~\ref{tab:table1}. 

\begin{table}
\caption{\label{tab:table1}Numerical values of the parameters used in calculations, adopted from Ref.~\cite{Clouet}.}
\begin{tabular}{lccccc}
 &\multicolumn{3}{c}{Zr}&\multicolumn{2}{c}{Sc}\\ \hline
 $T$ (K)&723&773&873
&723&773\\ 
 $c_{\rm{tot}}^{\rm{sat}}$ (\%)&0.029&0.055&0.159&0.019&0.039 \\
 $D$ (nm$^{2}$/s)&$0.232$&$3.144$&$235.54$&176.868&1134\\
\end{tabular}
\end{table}


In Fig.~\ref{Fig:2} we plot the chemical potential change on clusterization (per Al$_{3}$Zr formula unit) for the Al-Zr alloy as a function of the Zr volume fraction (bottom axis) or the nominal supersaturation (top axis) at $T=723~{\rm{K}}$, obtained by the cluster variation method (CVM) and the direct calculation method (DCM) from the MC simulations~\cite{Clouet}\footnote{In Ref.~\cite{Clouet} these values are calculated per lattice site. To rescale them per Al$_{3}$Zr formula unit, we multiply them by $4$.} together with the results of Eq.~(\ref{d_mu}), with the actual supersaturation calculated from ES~(\ref{EOS}). One can see that the calculated dependence is close to the CVM one and is practically identical to the DCM one, while the DS approximation is valid only in the low-supersaturation regime.

In Fig.~\ref{Fig:3} we plot the cluster size distributions in the Al-Zr and Al-Sc alloys, obtained from the MC simulations~\cite{Clouet}, performed at different temperatures and nominal supersaturations, together with the results of Eq.~(\ref{g_J_a}), with the actual supersaturation calculated from ES~(\ref{EOS}). From Fig.~\ref{Fig:3} one can see that the present theory is able to reproduce the simulation data on the cluster distribution functions with a good accuracy in all cases except those corresponding to the highest supersaturations of Zr at 723~K in Fig.~\ref{Fig:3}~(a). For a possible explanation of this discrepancy see discussion of the nucleation rate data in Fig.~\ref{Fig:4} below.

In Fig.~\ref{Fig:4} we display the nucleation rates for the Al$_{3}$Zr and Al$_{3}$Sc clusters, obtained from the MC simulations~\cite{Clouet}, performed at different temperatures and nominal supersaturations, together with the results of Eq.~(\ref{J_nucl}), with the actual supersaturation given by  ES~(\ref{EOS}). The DS result with $S_{\rm{act}}=S_{\rm{nom}}$ is also given as a reference.
One can see that the present theory gives a quantitative agreement with the MC simulations for both systems at all temperatures and not very high nominal supersaturations. 
Fig.~\ref{Fig:4} shows a general tendency of divergence between the simulation and theoretical data with the increase of supersaturation. The same tendency can be observed for the cluster size distribution data in Fig.~\ref{Fig:3}~(a). A probable reason for this discrepancy is the effect of frustration~\cite{Lepinoux2006}, which increases with the increase of the solute volume fraction. For the sake of brevity, this effect is neglected here. One can see that the values of the nucleation rate, calculated in the DS approximation, considerably overestimate the simulation data.

\section{\label{Conclusions}Conclusions and outlook}

In conclusion, the CNT-based Frenkel's model of heterophase structure of supersaturated solutions is used to construct an algebraic equation of supersaturation of a binary metastable solution, coupling the values of the nominal and the actual supersaturation with a function of the specific interface energy and temperature. This equation is also valid for saturated and undersaturated solutions. Application of this equation is shown to result in a much better (compared to the dilute solution model) agreement with MC simulation~\cite{Clouet} data on nucleation of Al$_{3}$Zr and Al$_{3}$Sc in model binary Al alloys. This approach may be further advanced by taking into account the effect of frustration~\cite{Lepinoux2006}. 

\section*{Acknowledgments}
The author is grateful to Dr. A. Turkin for numerous discussions and commenting the manuscript. I thank an anonymous reviewer for a suggestion to use the exact form of Eq.~(\ref{g_J_a}) instead of the approximate one. This work has been funded by the National Academy of Science of Ukraine, grant \# X-4-4/2017.

\section*{References}


\begin{thebibliography}{99}

\bibitem{Lin} Chen Lin,  Yang Zhang, Jing J Liu, and Xue Z Wang,
J. Cryst. Growth \textbf{469}, 59-64 (2017).

\bibitem{Dhivya} R. Dhivya, R. Ezhil Vizhi, and D. Rajan Babu,
J. Cryst. Growth \textbf{468}, 84-87 (2017).

\bibitem{Binder} Kurt Binder and Peter Virnau,
J. Chem. Phys. \textbf{145}, 211701 (2016).

\bibitem{Greer} A. L. Greer,
J. Chem. Phys. \textbf{145}, 211704 (2016).

\bibitem{Warrier} Pramod Warrier, M. Naveed Khan, Vishal Srivastava, C. Mark Maupin, and Carolyn A. Koh,
J. Chem. Phys. \textbf{145}, 211705 (2016).

\bibitem{Bi} Yuanfei Bi, Anna Porras, and Tianshu Li,
J. Chem. Phys. \textbf{145}, 211909 (2016).

\bibitem{Lifanov} Yuri Lifanov, Bart Vorselaars, and David Quigley,
J. Chem. Phys. \textbf{145}, 211912 (2016).

\bibitem{Legg} Benjamin A. Legg and James J. De Yoreo,
J. Chem. Phys. \textbf{145}, 211921 (2016).

\bibitem{Peters} Baron Peters,
J. Cryst. Growth \textbf{317}, 79-83 (2011).

\bibitem{Mangere} M. Mangere, J. Nathoo, and A.E. Lewis,
J. Cryst. Growth \textbf{312}, 3178-3182 (2010).

\bibitem{Landau} L.D. Landau and E.M. Lifshitz, \textit{Statistical Physics. Part 1. 3rd edition} (Butterworth-Heinemann, Oxford, 1980).

\bibitem{Kashchiev} D. Kashchiev, \textit{Nucleation: Basic Theory with Applications} (Butterworth-Heinemann, Oxford, 2000).

\bibitem{Frenkel} J. Frenkel,
J. Chem. Phys. \textbf{7}, 538 (1939). 

\bibitem{B_T} A. A. Turkin and A. S. Bakai,
Problems of Atomic Science and Technology \textbf{\#3(2)}, 394 (2007).

\bibitem{Lepinoux2009} J. Lepinoux, 
Acta Materialia \textbf{57}, 1086 (2009).

\bibitem{Lepinoux2010} J. Lepinoux, 
Philosophical Magazine \textbf{90}, 3261 (2010).

\bibitem{S_O} R. V. Shapovalov and O. A. Osmayev,
Problems of Atomic Science and Technology \textbf{\#1}, 273 (2012).

\bibitem{Borisenko1} Oleksandr Borysenko,
Condensed Matter Physics \textbf{18}, 23603 (2015).

\bibitem{Borisenko2} Alexander Borisenko,
Phys. Rev. E \textbf{93}, 052807 (2016)


\bibitem{Clouet} Emmanuel Clouet, Maylise Nastar, and Christophe Sigli,
Phys. Rev. B \textbf{69}, 064109 (2004).


\bibitem{Lepinoux2006} J. Lepinoux, 
Philosophical Magazine \textbf{86}, 5053 (2006).


\end{thebibliography}

\end{document}